\documentclass[aip,jmp,graphicx,preprint]{revtex4-1}
\usepackage{amsthm,epsfig,graphicx,amsmath,amssymb}

\font\bbsmall=msbm10 at 8pt

\newtheorem{theorem}{Theorem}[section]
\newtheorem{lemma}{Lemma}[section]

\newtheorem{example}{Example}[section]

\begin{document}


\title{On the absence of absolutely continuous spectra for Schr\"{o}dinger
operators on radial tree graphs}



\author{Pavel Exner}
 \email{exner@ujf.cas.cz}
 \affiliation{Doppler Institute for Mathematical Physics and Applied
 Mathematics, \\ Czech Technical University, B\v{r}ehov\'{a} 7,
 11519 Prague}  
 \affiliation{Nuclear Physics Institute ASCR, 25068
 \v{R}e\v{z} near Prague, Czechia}

 \author{Ji\v{r}\'{\i} Lipovsk\'{y}}
 \email{lipovsky@ujf.cas.cz}
 \affiliation{Nuclear Physics Institute ASCR, 25068
 \v{R}e\v{z} near Prague, Czechia}
 \affiliation{Institute of Theoretical Physics, Faculty of Mathematics and
 Physics, \\ Charles University, V Hole\v{s}ovi\v{c}k\'ach 2, 18000
 Prague}  


\date{\today}

\begin{abstract}
\noindent The subject of the paper are Schr\"odinger operators on tree graphs which are radial having the branching number $b_n$ at all the vertices at the distance $t_n$ from the root. We consider a family of coupling conditions at the vertices characterized by $(b_n-1)^2+4$ real parameters. We prove that if the graph is sparse so that there is a subsequence of $\{t_{n+1}-t_n\}$ growing to infinity, in the absence of the potential the absolutely continuous spectrum is empty for a large subset of these vertex couplings, but on the the other hand, there are cases when the spectrum of such a Schr\"odinger operator can be purely absolutely continuous.
\end{abstract}

\pacs{}

\maketitle 

\section{Introduction}

Quantum graphs became an immensely popular subject in the last two decades not only because of their numerous practical applications \cite{AGA}, but also because they are a good laboratory to study properties of quantum systems. The core of the appeal is that they exhibit mixed dimensional properties being locally one-dimensional, as long as a single edge is concerned, but globally multidimensional of many different types.

A class which attracted a particular attention are the tree graphs. An important question concerns free propagation of a particle on such graphs, i.e. the absolutely continuous spectrum of the corresponding tree Hamiltonians. It is known, for instance, that the \emph{ac} spectral component can survive a weak disorder coming from edge length variation -- cf.~Ref.~\onlinecite{ASW} and references therein. On the other hand, it was demonstrated recently by Breuer and Frank \cite{BF} that the spectrum on radial \emph{sparse} graphs in which a subsequence of edge lengths tends to infinity is purely singular.

The last named result was derived for the simplest vertex coupling usually called Kirchhoff. In this paper we address ourselves the question how does the propagation on a radial tree graph depend on coupling at the vertices. The family we consider is large: out of the $(b_n+1)^2$ parameters admissible at a tree vertex with the branching number $b_n$ by the self-adjointness requirement we will discuss a $[(b_n-1)^2+4]$-parameter subset. We will demonstrate that for a large part of it the result of Breuer and Frank is preserved, however, there are cases of vertex couplings for which the spectrum of the corresponding Hamiltonian has an absolutely continuous component or even it is purely absolutely continuous.

The method we are going to use is based on the seminal observation of Solomyak and coauthors -- cf.~Ref.~\onlinecite{SS}, references therein and developments in the subsequent work \cite{HP, BF} -- which makes it possible to reduce the problem to study of a family of Schr\"odinger operators on halfline, in our case with suitable generalized point interactions. What is important is that of all the vertex coupling parameters all but four will show up only at the boundary condition at the halfline endpoint. In analogy with Ref.~\onlinecite{BF} we will combine such a decomposition with an appropriate modification of a theorem by Remling \cite{R1}. As a preliminary we will summarize in the next three sections needed facts about Schr\"odinger operators on metric trees and parametrizations of generalized point interactions. In Sec.~5 we will then derive the decomposition mentioned above and in Sec.~6 we modify Remling's theorem for our purposes, and in the final section we combine these results to state and prove our claims.

\section{Schr\"{o}dinger operators on tree graphs}\label{schrodinger-tree}

Basic notions of nonrelativistic quantum mechanics on graphs are
nowadays well know so we can recall them only very briefly making
reference, e.g., to Ref.~\onlinecite{ES, KS, Ku} and an extensive 
bibliography in
the proceedings volume \cite{AGA}. Given a metric graph $\Gamma$
we use $L^2(\Gamma)$ as the state Hilbert space. The Hamiltonian
acts as a one-dimensional Schr\"odinger operator on each edge; in
the particular case when there is no potential it is simply
$f_j\mapsto -f''_j$ on the $j$-th edge. To make this operator
self-adjoint suitable coupling conditions have to be imposed at
the vertices. The simplest one are \emph{free} conditions (often
also called Kirchhoff) which require function continuity at the
vertex together with vanishing sum of the derivatives. Below we
will introduce a wide family of other coupling conditions we are
going to consider in this paper.

By a seminal observation of Sobolev and Solomyak~\cite{SS}
a~Schr\"{o}dinger operator on a~homogeneous rooted tree graph with
free coupling conditions at the vertices is unitarily equivalent
to the orthogonal sum of operators acting on $L^2(\mathbb{R}_+)$,
namely one-dimensional Schr\"odinger operators with appropriate
singular interactions. We are going to discuss how this result
generalizes to a~larger class of coupling conditions, branching
numbers, and different lengths of the edges under the assumption
that the potential $V(|x|)$ is real, bounded and measurable
depending on the distance from the root $|x|$ only. This
equivalence will be subsequently our main technical tool to
demonstrate claims about absolutely continuous spectrum of
Schr\"odinger operators on such trees.

Speaking about tree graphs, we will use a notation similar
to that of Ref.~\onlinecite{SS,EFK,HP}. Let
$\Gamma$ be a~rooted metric tree graph with the root labeled by
$o$. We denote by $|x|$ the distance between the point $x$ of the
graph and the root $o$. The {\em branching number} $b(v)$ of the
vertex $v$ is the number of vertices emanating from this vertex
``forward'', i.e. the vertex $v$ connects one edge of the previous
generation with $b(v)$ outgoing edges. In this sense,
 $b(o) = 1$, while for the other vertices we assume $b(v)\geq 1$.

\begin{figure}
   \begin{center}
     \includegraphics{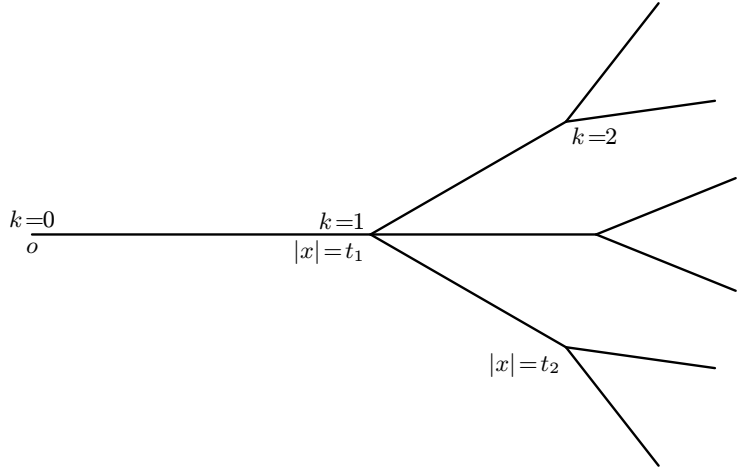}
     \caption{An example of a~radial tree for $b_0 = 1$, $b_1=3$, $b_2=2$}\label{figtree}
   \end{center}
\end{figure}

We say that the vertex $v$ of a~tree graph $\Gamma$ belongs to the
$k$-th generation if there are just $k-1$ vertices on the shortest
path between $v$ and $o$. We write $\mathrm{gen\,} v = k$, where
$k$ is a natural number or zero which is by definition associated
with the root. We call the tree graph {\em radial} if the
branching numbers for all the vertices of the same generation are
equal and the edges emanating from these vertices have equal
lengths (cf.~Fig.~\ref{figtree}). For radial graphs we introduce
$t_k$ as the distance between the root and the vertices in the
$k$-th generation, and $b_k$ as the branching number of the $k$-th
generation vertices; for the root we put $b_0 = 1$ and $t_0 = 0$.
Furthermore, one defines the {\em branching function} $g_0(t) :
{\mathbb{R}_+\to \mathbb{N}}$ by
 $$
    g_0 (t) := b_0 b_1 \dots b_k \quad \hbox{for}\quad t \in (t_{k},t_{k+1})\,.
 $$
The tree graph is called {\em homogeneous} if the branching number
$b$ for all vertices except of $o$ is the same.

Vertices of a tree graph are naturally ordered. We say that vertex
$w$ succeeds vertex $v$, or $w\succeq v$, if $v$ lies on the
shortest path from $o$ to $w$; we also say that $v$ precedes $w$.
Notice that the ordering relation $\succeq$ is reflexive, i.e. a
vertex precedes and succeeds itself, and that the ordering
naturally extends to edges. Furthermore, one defines the {\em
vertex subtree} $\Gamma_{\succeq v}$ as the set of vertices and
edges succeeding $v$, and the {\em edge subtree} $\Gamma_{\succeq
e}$ as the union of the edge $e$ and the vertex subtree
corresponding to its vertex remoter from the origin.

To construct the decomposition mentioned above we need means to
characterize permutation properties of graph edges. Consider
a~radial tree graph with the vertex $v$ of the $k$-th generation;
since $v$ is fixed we for simplicity write $b \equiv b_k$. We
denote the edges emanating from $v$ by $e_j, j\in \{1,\dots, b\}$.
Consider next the operator $Q_v$ on $L^{2}(\Gamma_{\succeq v})$
which cyclically shifts indices of the functions $f_j$ on edge
subtrees $\Gamma_{\succeq {e_j}}$ in the following way,
 $$
   Q_v : f_j \mapsto f_{j+1}\,,
 $$
where we have identified $f_{b+1}$ with $f_1$; each $f_j$ is
naturally a collection of functions referring to the edges
succeeding $e_j$. Since $Q_v^{b}=\mathrm{id}$, the operator has
eigenvalues ${\rm e}^{2\pi is/b}, s\in \{0,\dots,b-1\}$. We denote
the corresponding eigenspaces by $L^{2}_s (\Gamma_{\succeq
v}):=\mathrm{Ker}(Q_v - {\rm e}^{2\pi is/b}\, \mathrm{id})$. We
call the function $f\in L^{2}(\Gamma_{\succeq v})$ {\em s-radial
at the vertex $v$} if $f \in L^{2}_s (\Gamma_{\succeq v})$ and $f
\in L^{2}_0 (\Gamma_{\succeq v'})$ holds for all vertices $v'$
succeeding $v$. The set of all such functions we denote by
$L^{2}_{s, \mathrm{rad}} (\Gamma_{\succeq v})$. In particular, the
0-radial functions will be simply called {\em radial}.

Now we can pass to the coupling conditions needed to make the
Hamiltonian self-adjoint. As usual we restrict our attention to
the \emph{local} ones, i.e. those coupling boundary values in each
particular vertex separately. In general, admissible couplings at
a vertex $v$ can be characterized by $(b_k+1)^2$ real parameters,
or equivalently, by a unitary $[(b_k+1)]\times [(b_k+1)]$ matrix
\cite{Ha,KS}. In order to construct the unitary equivalence with
halfline problems mentioned above, we have to restrict our
consideration to a~$[(b_k -1)^2+4]$-parameter subset by adopting
the assumption that all the emanating edges are equivalent.
Moreover, the unitary equivalence requires the parameters of the
coupling to be equal for all the vertices of the same generation.
Later we will show that only some of these parameters influence
the spectrum as a set.

To be specific, at a~vertex $v$ belonging to the $k$-th
generation, $k\geq 1$, we impose following coupling conditions
 \begin{eqnarray}
    \sum_{j=1}^{b_k} f_{vj+}'- f_{v-}'& = & \frac{\alpha_{\mathrm{t}k}}{2} \left(\frac{1}{b_k}\sum_{j=1}^{b_k}f_{vj+}+ f_{v-}\right)+
    \frac{\gamma_{\mathrm{t}k}}{2}\left(\sum_{j=1}^{b_k} f_{vj+}'+ f_{v-}'\right),\label{3-fcoupl1}
\\
    \frac{1}{b_k}\sum_{j=1}^{b_k}f_{vj+}- f_{v-}& = & -\frac{\bar\gamma_{\mathrm{t}k}}{2} \left(\frac{1}{b_k}\sum_{j=1}^{b_k}f_{vj+}+ f_{v-}\right)+
    \frac{\beta_{\mathrm{t}k}}{2}\left(\sum_{j=1}^{b_k} f_{vj+}'+ f_{v-}'\right).\label{3-fcoupl2}
     \end{eqnarray}
 \begin{equation}
    (U_k - I) V_k\Psi_v + i (U_k + I) V_k\Psi_v' = 0\,,\label{3-fcoupl3}
 \end{equation}
where the index $j$ distinguishes the edges emanating from $v$,
the subscript minus refers to the ingoing (or preceding) edge, and
 \begin{eqnarray*}
    \Psi_v &\!:=\!& (f_{v1+},f_{v2+},\dots,f_{vb_k +})^\mathrm{T},
    \\ [.5em]
    \Psi_v' &\!:=\!& (f_{v1+}',f_{v2+}',\dots,f_{vb_k +}')^\mathrm{T}
 \end{eqnarray*}
As indicated above the coefficients $\alpha_{\mathrm{t}k},
\beta_{\mathrm{t}k}\in \mathbb{R}$, and $\gamma_{\mathrm{t}k}\in
\mathbb{C}$ are the same for all the vertices belonging to the
$k$-th generation. The subscript t indicates that they describe
the coupling on the tree graph and we will use it in order to
avoid confusion with the halfline counterpart in the following
sections. Coupling between vectors $\Psi_v$ and $\Psi_v'$ is
described by a $(b_k-1) \times (b_k-1)$ unitary matrix $U_k$,
while $V_k$ stands for an arbitrary $b_k \times (b_k-1)$ matrix
with orthonormal rows which all are perpendicular to the vector
$(1,1,\dots,1)$. In other words,$V_k$ is the $(b_k-1)$-dimensional
projection to the orthogonal complement of $(1,1,\dots,1)$, and
the vectors $V_k (f_1 (\cdot), \dots, f_{b_k} (\cdot))^\mathrm{T}$
form an orthonormal basis in $L^{2} (\Gamma_{\succeq v})\ominus
L^{2}_{0, \mathrm{rad}} (\Gamma_{\succeq v})$; here again $f_j$
stands for a collection of functions on the appropriate edge
subgraph. The same coupling conditions are applied to all vertices
in the same generation, i.e. neither $U_k$ nor $V_k$ depends on
the particular $k$-th generation vertex at which they are applied.

To have the Hamiltonian well defined we have to fix also the
boundary condition at the tree root. We choose them in the Robin
form,
 \begin{equation}
    f_o ' + f_o \,\tan\frac{\theta_0}{2}  = 0\,,
    \quad \theta_0 \in (\pi/2,\pi/2] \,. \label{3-fcoupl4}
 \end{equation}

Let us denote by $\mathbf{H}$ the Hamiltonian acting as
$-\mathrm{d}^2/\mathrm{d} x^2 + V(|x|)$ on a~radial tree graph
$\Gamma$ with the branching numbers $b_k$ described above and the
potential depending on the distance from the root only. We will
suppose that the potential is essentially bounded, $V\in
L^\infty(\Gamma)$; this assumption is done for the sake of
simplicity only and can easily be weakened.

The domain of this operator consists then of functions
$f(x)\in\sum_{e\in \Gamma} \oplus H^{2}(e)$ satisfying the
coupling conditions (\ref{3-fcoupl1})--(\ref{3-fcoupl4}). In the
following, the Hamiltonian on the tree graph is denoted by a bold
$\mathbf{H}$ while the corresponding Hamiltonians of its halfline
counterparts are denoted by $H$.

 \begin{lemma}\label{sa}
 The above differential expression together with the coupling
 conditions (\ref{3-fcoupl1})--(\ref{3-fcoupl4}) define a self-adjoint
 operator.
 \end{lemma}
 \begin{proof}
 The coupling (\ref{3-fcoupl1})--(\ref{3-fcoupl3}) can be concisely
 expressed by the equation
 $$
   A_v \begin{pmatrix}f_{v-}\\\Psi_{v}\end{pmatrix}+ B_v \begin{pmatrix}-f_{v-}'\\\Psi_{v}'\end{pmatrix}=0\,,
 $$
where
  \begin{eqnarray*}
   A_v &\!:=\!& \begin{pmatrix}
   -\frac{\alpha_{\mathrm{t}k}}{2}&
   -\frac{1}{b_k}\frac{\alpha_{\mathrm{t}k}}{2}&
   -\frac{1}{b_k}\frac{\alpha_{\mathrm{t}k}}{2}&\dots&
   -\frac{1}{b_k}\frac{\alpha_{\mathrm{t}k}}{2} \\[.3em]
   -(1-\frac{\bar\gamma_{\mathrm{t}k}}{2})&
   \frac{1}{b_k}(1+\frac{\bar\gamma_{\mathrm{t}k}}{2})&
   \frac{1}{b_k}(1+\frac{\bar\gamma_{\mathrm{t}k}}{2})&
   \dots&\frac{1}{b_k}(1+\frac{\bar\gamma_{\mathrm{t}k}}{2}) \\[.3em]
   0&\multicolumn{4}{c}{(U_k -I)V_k}
   \end{pmatrix}\,,
 \\ [.5em]
   B_v &\!:=\!& \begin{pmatrix}
   (1+\frac{\gamma_{\mathrm{t}k}}{2})&
   1-\frac{\gamma_{\mathrm{t}k}}{2}&
   1-\frac{\gamma_{\mathrm{t}k}}{2}&\dots&
   1-\frac{\gamma_{\mathrm{t}k}}{2} \\[.3em]
   \frac{\beta_{\mathrm{t}k}}{2}&-\frac{\beta_{\mathrm{t}k}}{2}&
   -\frac{\beta_{\mathrm{t}k}}{2}&\dots&-\frac{\beta_{\mathrm{t}k}}{2} \\[.3em]
   0&\multicolumn{4}{c}{i(U_k +I)V_k}
  \end{pmatrix} \,.
  \end{eqnarray*}

\noindent Using the standard condition form of Kostrykin and
Schrader \cite{KS} we need to check hermiticity of matrix $A_v
B_v^*$. A simple calculation yields
 $$
   A_v B_v^* = \begin{pmatrix}
   -\alpha_{\mathrm{t}k}&0&0\\
   0&-\beta_{\mathrm{t}k}&0\\
   0&0&-i(U_k -I)(U_k^*+I)
   \end{pmatrix} = B_v A_v^*\,.
 $$
We have used here the projection property of the matrix $V_k$,
i.e. $V_k V_k^* = I$, and unitarity of the matrix $U_k$, i.e. $U_k
U_k^* = I$ which gives $-i(U_k -I)(U_k^*+I) = i(U_k +I)(U_k^*-I)$.

Furthermore, one needs to check that the rectangular matrix $(A_v,B_v)$
has maximal rank. To make its first two rows linearly dependent, one
has to satisfy simultaneously the conditions $C \alpha_{\mathrm{t}k} = 2 - \bar \gamma_{\mathrm{t}k}$ and $-C \alpha_{\mathrm{t}k} = 2 + \bar \gamma_{\mathrm{t}k}$ for some constant $C$, and similar conditions for $\beta_{\mathrm{t}k}$; this leads to a contradiction. Linear dependence of the first and the $i$-th row, $i>2$, requires first that $\alpha_{\mathrm{t}k}$ vanishes, using this fact we further get $\sum_j (u_{ij}+\delta_{ij})v_{jm} = C$ for entries of the matrices $U_k$ and $V_k$, and similarly $\sum_j (u_{ij}+\delta_{ij})v_{jm} = C$ for
all $m$. Hence $2 v_{jm} = C$ should hold for all $m$, however, $V_k$ has rows perpendicular to $(1,\dots,1)$, which is again a contradiction. The
same argument applies to the second and the $i$-th row, $i>2$. Finally, to make the $i$-th and $j$-th row, $i,j>2$, linearly dependent, the conditions $\sum_m (u_{im}+\delta_{im})v_{mn} = C(u_{jm}+\delta_{jm})v_{mn}$ and $\sum_m (u_{im}-\delta_{im})v_{mn} = C(u_{jm}-\delta_{jm})v_{mn}$ must be
satisfied for some $C$, which amounts to linear dependence of $i$-th and $j$-th row of $V_k$; in that way have managed to reduce the assumption \emph{ad absurdum}. It is easy to check the selfadjointness condition for the root.
 \end{proof}

\newpage
\section{Parameterizations of generalized point \\ interactions}

There are multiple ways to describe the four-parameter generalized point interaction (GPI) on the line which can be regarded as a simple graph with a single vertex connection two semiinfinite leads. Before proceeding with the construction of the unitary equivalence between the Hamiltonian on
a graph and a direct sum of halfline operators, let us summarize some known results. As a graph vertex coupling, of course, the GPI can be described by the standard coupling conditions mentioned above \cite{KS} or one of their unique forms \cite{Ha, CET}. We will recall two other descriptions which leave out some GPI's becoming singular for certain values of the parameters but have other advantages: the first one coming from Ref.~\onlinecite{EG} includes the important particular cases of $\delta$ and $\delta'$ interactions in a symmetric way, the other is most commonly used in this context.

For brevity, we label the limits of functional value and the derivative from the right by $y_{+}$ and $y_{+}'$, respectively, and analogously for the functional value and derivative from the left. The first of the above mentioned parameterizations,
 \begin{eqnarray}
    y_{+}'- y_{-}'&=& \frac{\alpha}{2} (y_{+}+ y_{-})+\frac{\gamma}{2}(y_{+}'+ y_{-}')\,,\label{3-a1}\\
    y_{+}- y_{-}&=& -\frac{\bar\gamma}{2} (y_{+}+ y_{-})+\frac{\beta}{2}(y_{+}'+ y_{-}')\label{3-a2}
 \end{eqnarray}
is characterized by a~matrix $\mathcal{A}=\begin{pmatrix}\alpha&\gamma\\ -\bar\gamma&\beta\end{pmatrix}$ with $\alpha, \beta \in \mathbb{R}$ and $\gamma \in \mathbb{C}$. While it is not universal, this parameterization describes almost all selfadjoint extensions of the operator $-\mathrm{d}^2/\mathrm{d}x^2$ restricted to the subspace $\{f\in H^{2}(\mathbb{R}):y_+=y_+'=y_-=y_-'=0\}$, the exceptions being separated halflines with Dirichlet or Neumann imposed on both sides. The form (\ref{3-a1})--(\ref{3-a2}) reduces to the $\delta$-condition case of strength $\alpha$ if $\beta = \gamma = 0$, and to the $\delta'$-condition case of strength $\beta$ if $\alpha = \gamma = 0$. The second parametrization to consider is
 \begin{equation}
    \begin{pmatrix}y_{+}'\\-y_{-}'\end{pmatrix} = \begin{pmatrix}a&c\\\bar c&d\end{pmatrix}
        \begin{pmatrix}y_{+}\\y_{-}\end{pmatrix}\,,\label{3-b}
 \end{equation}
with $a, d \in \mathbb{R}$ and $c \in \mathbb{C}$. This parametrization decouples the two leads if $c =0$.

Recall first how to pass from (\ref{3-a1})--(\ref{3-a2}) to (\ref{3-b}). We rewrite the former as
 $$
    \begin{pmatrix}1-\frac{\gamma}{2}&1+\frac{\gamma}{2}\\[.2em]
    -\frac{\beta}{2}&\frac{\beta}{2} \end{pmatrix}
        \begin{pmatrix}y_{+}'\\-y_{-}'\end{pmatrix} =
        \begin{pmatrix}\frac{\alpha}{2}&\frac{\alpha}{2}\\[.2em]
        -1-\frac{\bar\gamma}{2} &1-\frac{\bar\gamma}{2}\end{pmatrix}
        \begin{pmatrix}y_{+}\\y_{-}\end{pmatrix}\,,
 $$
and a simple calculation yields
 \begin{multline*}
     \begin{pmatrix}a&c\\\bar c&d\end{pmatrix} =
        \begin{pmatrix}1-\frac{\gamma}{2}&1+\frac{\gamma}{2}\\-\frac{\beta}{2}&\frac{\beta}{2} \end{pmatrix}^{-1}
        \begin{pmatrix}\frac{\alpha}{2}&\frac{\alpha}{2}\\-1-\frac{\bar\gamma}{2} &1-\frac{\bar\gamma}{2}\end{pmatrix}
 \\
    =\frac{1}{4\beta}\begin{pmatrix}4+\mathrm{det}\mathcal{A}+4\,\mathrm{Re\,}\gamma&
        -4+\mathrm{det}\mathcal{A}-4i\,\mathrm{Im\,}\gamma\\-4+\mathrm{det}\mathcal{A}+4i\,\mathrm{Im\,}\gamma&
        4+\mathrm{det}\mathcal{A}-4\,\mathrm{Re\,}\gamma\end{pmatrix};
 \end{multline*}
notice that in view of $\beta$ in the denominator the parametrization (\ref{3-b}) does not contain the case of $\delta$-interaction. Conversely, to pass from (\ref{3-b}) to (\ref{3-a1})--(\ref{3-a2}) it is convenient to introduce another basis,
 \begin{eqnarray*}
    g_1 = y_+ + y_-\,,\quad g_2= y_+' + y_-'\,,\\
    g_3 = y_+ - y_-\,,\quad g_4= y_+' - y_-'\,.
 \end{eqnarray*}
Expressing $y_\pm$ and $y'_\pm$ from here,
one can rewrite the equation (\ref{3-b}) as
 $$
   \begin{pmatrix}1&c-a\\1&d-\bar c\end{pmatrix}\begin{pmatrix}g_4\\g_3\end{pmatrix} =
    \begin{pmatrix}a+c &-1\\d+\bar c& 1\end{pmatrix}\begin{pmatrix}g_1\\ g_2\end{pmatrix}
 $$
and therefore
 $$
   \begin{pmatrix}\alpha&\gamma\\-\bar\gamma&\beta\end{pmatrix} = 2\,
    \begin{pmatrix}1&c-a\\1&d-\bar c\end{pmatrix}^{-1}\begin{pmatrix}a+c&- 1\\d+\bar c &1\end{pmatrix},
 $$
so after another simple calculation we can summarize the relations as follows.
 \begin{lemma}\label{3-lemma_prevod}
The correspondence of the GPI coupling conditions (\ref{3-a1})--(\ref{3-a2}) and (\ref{3-b}) is given by
 \begin{eqnarray*}
   \begin{pmatrix}a&c\\\bar c&d\end{pmatrix} &=&
     \frac{1}{4\beta}\begin{pmatrix}4+\mathrm{det}\mathcal{A}+4\,\mathrm{Re\,}\gamma&
        -4+\mathrm{det}\mathcal{A}-4i\,\mathrm{Im\,}\gamma\\-4+\mathrm{det}\mathcal{A}+4i\,\mathrm{Im\,}\gamma&
        4+\mathrm{det}\mathcal{A}-4\,\mathrm{Re\,}\gamma\end{pmatrix}\,,\\[.5em]
   \begin{pmatrix}\alpha&\gamma\\-\bar\gamma&\beta\end{pmatrix}
   &=&\frac{4}{a+d-2\,\mathrm{Re\,}c}\begin{pmatrix}ad-|c|^2&\frac{1}{2}(a-d)-i \,\mathrm{Im\,}c\\
    -\frac{1}{2}(a-d)-i\,\mathrm{Im\,}c &1\end{pmatrix}\,.
 \end{eqnarray*}
 \end{lemma}

 Let us also recall that the universal parametrization of a GPI  according to Ref.~\onlinecite{Ha} using  $2\times 2$ unitary matrices~$U$,
 $$
   (U-I)\begin{pmatrix}y_+\\y_-\end{pmatrix}+i(U+I)\begin{pmatrix}y_+'\\-y_-'\end{pmatrix}=0
 $$
where $U = \mathrm{e}^{i\xi}\begin{pmatrix}u_1&u_2\\-\bar u_2&\bar u_1\end{pmatrix}$
with $u_1, u_2 \in \mathbb{C}$, $|u_1|^2 + |u_2|^2 = 1$ and $\xi \in [0,\pi)$ can be according to Ref.~\onlinecite{EF} related to  the parametrization (\ref{3-a1})--(\ref{3-a2}) by
 \begin{eqnarray*}
   u_1 &=& \frac{-2(\alpha+\beta)+4i\mathrm{\,Re\,}\gamma}
    {\sqrt{(\alpha\beta+|\gamma|^2)^2+4\alpha^2+4\beta^2+8|\gamma|^2+16}}\,,\\
   u_2 &=& \frac{1}{2i}\frac{\alpha\beta+|\gamma|^2-4-4i\mathrm{\,Im\,}\gamma}
    {\sqrt{(\alpha\beta+|\gamma|^2)^2+4\alpha^2+4\beta^2+8|\gamma|^2+16}}\,,\\
   \tan{\xi} &=& \frac{\alpha\beta+|\gamma|^2+4}{2(\alpha-\beta)}\,.
\end{eqnarray*}

\section{Mapping to a halfline}\label{sec_mapping}

As indicated our goal is to map the tree problem unitarily to a family of halflines. In this section, we will look at it ``locally'' investigating  which halfline coupling conditions can correspond to (\ref{3-fcoupl1})--(\ref{3-fcoupl4}). Recall that the main idea of
the unitary equivalence employed in Ref.~\onlinecite{SS, HP, EFK} consists of  identification of ``symmetric'' functions,  $f\in L^2_{0,\mathrm{rad}}(\Gamma)$, with the corresponding function on the halfline. This is achieved through the isometry $\Pi : f \to \varphi$, $\varphi(t) = f(x)$ for $t =|x|$ of  $L^2_{0,\mathrm{rad}}(\Gamma)$
into the weighted space $L^2(\mathbb{R}_+,g_0)$ with the norm
 $$
    \|\varphi \|^2_{L^2(\mathbb{R}_+,g_0)} =
        \int_{\mathbb{R}_+}|\varphi(t)|^2 g_0(t) \,\mathrm{d}t
 $$
combined with passing to $L^2(\mathbb{R})$ by the isometry $y(t):=g_0^{1/2}(t) \varphi(t)$ and the relations
 \begin{eqnarray*}
      y_{k+} &=& (b_0\cdot\ldots\cdot b_k)^{1/2} \varphi_{k+}\,,\\
      y_{k-} &=& (b_0\cdot\ldots\cdot b_{k-1})^{1/2} \varphi_{k-}\,,
 \end{eqnarray*}
for the boundary values at the vertices.

We can substitute the last relations into (\ref{3-fcoupl1})--(\ref{3-fcoupl4}) and
divide both sides of these four equations by $(b_0 \dots b_{k-1})^{-1/2}$.
In view of the linearity of the coupling conditions (\ref{3-fcoupl1})--(\ref{3-fcoupl4})
the passage from $f(x)$ to $y(t)$ is for a vertex of the $k$-th generation equivalent to the replacements
 \begin{eqnarray}
   f_{v-} \to y_{k-}\,,&\quad& f_{v-}' \to y_{k-}'\,,\label{3-change1}\\
   \frac{1}{b_k}\sum_{j =1}^{b_k} f_{vj+} \to b_{k}^{-1/2}y_{k+}\,,&\quad& \sum_{j=1}^{b_k}f_{jv+}' \to b_{k}^{1/2}y_{k+}'\label{3-change2}\,.
 \end{eqnarray}
Since rearrangement of equations (\ref{3-fcoupl1})--(\ref{3-fcoupl2}) after substitutions (\ref{3-change1})--(\ref{3-change2}) into the form (\ref{3-a1})--(\ref{3-a2}) is more complicated, we first investigate the change of the coupling
 \begin{equation}
    \begin{pmatrix}\sum_{j=1}^{b}f_{j+}'\\-f_{-}'\end{pmatrix} =
    \begin{pmatrix}a_{\mathrm{t}}&c_{\mathrm{t}}\\\bar c_{\mathrm{t}}&d_{\mathrm{t}}\end{pmatrix}
    \begin{pmatrix}\frac{1}{b}\sum_{j=1}^{b}f_{j+}\\f_{-}\end{pmatrix}\,,
 \end{equation}
which corresponds to the parametrization (\ref{3-b}). For simplicity, we have omitted here the indices $v$ and $k$. Using (\ref{3-change1})--(\ref{3-change2}) one obtains
 \begin{equation*}
    \begin{pmatrix}y_{+}'\\-y_{-}'\end{pmatrix} =
    \begin{pmatrix}b^{-1}a_{\mathrm{t}}&b^{-1/2}\,c_{\mathrm{t}}\\
    b^{-1/2}\,\bar c_{\mathrm{t}}&d_{\mathrm{t}}\end{pmatrix}
    \begin{pmatrix}y_{+}\\y_{-}\end{pmatrix}\,,
 \end{equation*}
thus the appropriate coupling parameters for the halfline are
 \begin{equation*}
   a_{\mathrm{h}}=b^{-1}a_{\mathrm{t}}\,,\quad c_{\mathrm{h}}=b^{-1/2}c_{\mathrm{t}}\,,
    \quad d_{\mathrm{h}}=d_{\mathrm{t}}\,.
 \end{equation*}
The condition (\ref{3-fcoupl3}) is for $f\in L^2_{0,\mathrm{rad}}(\Gamma)$ satisfied trivially and the root condition (\ref{3-fcoupl4}) is not affected by considered transformation.

Now we can employ Lemma~\ref{3-lemma_prevod} to find the correspondence of the coupling parameters in (\ref{3-fcoupl1})--(\ref{3-fcoupl2}) and those of (\ref{3-a1})--(\ref{3-a2}) on the halfline. If $\beta_{\mathrm{t}} \not = 0$ we have
 $$
   \alpha_{\mathrm{h}}= \frac{4 b^{-1}(a_{\mathrm{t}}d_{\mathrm{t}} -
    |c_{\mathrm{t}}|^2)}{b^{-1}a_{\mathrm{t}}+d_{\mathrm{t}}-2b^{-1/2}\mathrm{\,Re\,}c_{\mathrm{t}}}
   = \frac{16 \alpha_{\mathrm{t}}}{4(b^{1/2}+1)^2+\mathrm{det}\mathcal{A_{\mathrm{t}}}
    (b^{1/2}-1)^2+4(1-b)\mathrm{\,Re\,}\gamma_{\mathrm{t}}}\,,
 $$
and similarly
 \begin{eqnarray*}
   \beta_{\mathrm{h}} &\!=\!& \frac{16\, b\, \beta_{\mathrm{t}}}{4(b^{1/2}+1)^2+\mathrm{det}\mathcal{A_{\mathrm{t}}}
    (b^{1/2}-1)^2+4(1-b)\mathrm{\,Re\,}\gamma_{\mathrm{t}}}\,,
  \\[.5em]
   \gamma_{\mathrm{h}}&\!=\!&2\: \frac{(1-b)(4+\mathrm{det}\mathcal{A_{\mathrm{t}}})+8ib^{1/2}\mathrm{\,Im\,}\gamma_{\mathrm{t}}
    +4(b+1)\mathrm{\,Re\,}\gamma_{\mathrm{t}}}{4(b^{1/2}+1)^2+\mathrm{det}\mathcal{A_{\mathrm{t}}}
    (b^{1/2}-1)^2+4(1-b)\mathrm{\,Re\,}\gamma_{\mathrm{t}}}\,.
 \end{eqnarray*}
In the remaining case $\beta_{\mathrm{t}} = 0$ we use the basis $g_i$, $i =1,\dots,4$ introduced in previous section. The transformation (\ref{3-change1})--(\ref{3-change2}) then becomes
 \begin{eqnarray*}
   g_1 \to \frac{b^{-1/2}+1}{2}\tilde g_1 + \frac{b^{-1/2}-1}{2}\tilde g_3\,,\quad
   g_2 \to \frac{b^{1/2}+1}{2}\tilde g_2 + \frac{b^{1/2}-1}{2}\tilde g_4\,,\\
   g_3 \to \frac{b^{-1/2}+1}{2}\tilde g_3 + \frac{b^{-1/2}-1}{2}\tilde g_1\,,\quad
   g_4 \to \frac{b^{1/2}+1}{2}\tilde g_4 + \frac{b^{1/2}-1}{2}\tilde g_2\,.
 \end{eqnarray*}
Substituting from here into the coupling conditions
 $$
   \begin{pmatrix} g_4\\ g_3\end{pmatrix} =
    \frac{1}{2}\begin{pmatrix}\alpha_{\mathrm{t}}&\gamma_{\mathrm{t}}\\
    -\bar\gamma_{\mathrm{t}}&\beta_{\mathrm{t}}\end{pmatrix}
    \begin{pmatrix} g_1\\ g_2\end{pmatrix}
 $$
we get a pair of equations. From the second of them one obtains
 $$
   \gamma_{\mathrm{h}} = 2\:\frac{2(b^{-1/2}-1)+
	 \gamma_{\mathrm{t}}(b^{-1/2}+1)}{2(b^{-1/2}+1)+\gamma_{\mathrm{t}}(b^{-1/2}-1)}
\\
   =2\: \frac{(1-b)(4+|\gamma_{\mathrm{t}}|^2)+8ib^{1/2}\mathrm{\,Im\,}\gamma_{\mathrm{t}}
	 +4(b+1)\mathrm{\,Re\,}\gamma_{\mathrm{t}}}{4(b^{1/2}+1)^2+|\gamma_{\mathrm{t}}|^2
	(b^{1/2}-1)^2+4(1-b)\mathrm{\,Re\,}\gamma_{\mathrm{t}}}\,.
 $$
and subsequently, substituting $\tilde g_3 = -\frac{1}{2}\bar \gamma_{\mathrm{h}} \tilde g_1$ into the first one we get
 $$
   \alpha_{\mathrm{h}}
   = \frac{16 \alpha_{\mathrm{t}}}{4(b^{1/2}+1)^2+|\gamma_{\mathrm{t}}|^2
    (b^{1/2}-1)^2+4(1-b)\mathrm{\,Re\,}\gamma_{\mathrm{t}}}\,.
 $$
It holds trivially $\beta_{\mathrm{h}}=0$, and therefore, the expressions computed for $\beta_{\mathrm{t}}\not =0$ using Lemma~\ref{3-lemma_prevod} can be used also for $\beta_{\mathrm{t}} =0$ as well.

To list the remaining situations, Dirichlet or Neumann conditions obviously do not change under the transformation (\ref{3-change1})--(\ref{3-change2}) since $f_+ = f_- = 0$ implies $y_+ = y_- = 0$ and $\sum_{j=1}^b f_{j+}' = f_-' = 0 \quad\Rightarrow\quad  y_+' = y_-' = 0$. Finally, if the denominator in the above expression vanishes, $\gamma_{\mathrm{t}} = 2\frac{b^{1/2}+1}{b^{1/2}-1}$, one has two subcases, $\alpha_{\mathrm{t}} = 0$, $\beta_{\mathrm{t}} \not = 0$ and $\alpha_{\mathrm{t}} \not = 0$, $\beta_{\mathrm{t}} = 0$. Let us summarize the results of the above considerations.
 \begin{lemma}\label{3-lemma_th}
The vertex coupling conditions (\ref{3-fcoupl1})--(\ref{3-fcoupl4}) change under the transformation (\ref{3-change1}) -- (\ref{3-change2}) into
 \begin{eqnarray}
    y_{k+}'- y_{k-}'&\!=\!& \frac{\alpha_{\mathrm{h}k}}{2} (y_{k+}+ y_{k-})+
    \frac{\gamma_{\mathrm{h}k}}{2}(y_{k+}'+ y_{k-}')\,, \label{ycoupl1} \\
    y_{k+}- y_{k-}&\!=\!& -\frac{\bar\gamma_{\mathrm{h}k}}{2} (y_{k+}+ y_{k-})+
    \frac{\beta_{\mathrm{h}k}}{2}(y_{k+}'+ y_{k-}')\,, \label{ycoupl2}\\
    y(0+)' &\!+\!& y(0+) \,\tan\frac{\theta_0}{2}  = 0\,,\label{ycoupl3}
 \end{eqnarray}
where
 \begin{eqnarray}
   \alpha_{\mathrm{h}k}
   &\!:=\!& \frac{16 \alpha_{\mathrm{t}k}}{4(b_k^{1/2}+1)^2+\mathrm{det}\mathcal{A}_{\mathrm{t}k}
    (b_k^{1/2}-1)^2+4(1-b_k)\mathrm{\,Re\,}\gamma_{\mathrm{t}k}}\,, \label{alphah} \\
   \beta_{\mathrm{h}k}&\!:=\!& \frac{16\, b_k\, \beta_{\mathrm{t}k}}{4(b_k^{1/2}+1)^2+\mathrm{det}\mathcal{A}_{\mathrm{t}k}
    (b_k^{1/2}-1)^2+4(1-b_k)\mathrm{\,Re\,}\gamma_{\mathrm{t}k}}\,, \label{betah} \\
   \gamma_{\mathrm{h}k}&\!:=\!& 2\,\cdot\, \frac{(1-b_k)(4+\mathrm{det}\mathcal{A}_{\mathrm{t}k})+
    8ib_k^{1/2}\mathrm{\,Im\,}\gamma_{\mathrm{t}k}
    +4(b_k+1)\mathrm{\,Re\,}\gamma_{\mathrm{t}k}}{4(b_k^{1/2}+1)^2+\mathrm{det}\mathcal{A}_{\mathrm{t}k}
    (b_k^{1/2}-1)^2+4(1-b_k)\mathrm{\,Re\,}\gamma_{\mathrm{t}k}}\,. \nonumber
 \end{eqnarray}
The conditions $f_{v+} = f_{v-} = 0$ or $\sum_{j=1}^{b_k} f_{vj+}' = f_{v-}' = 0$ transform similarly to $y_{k+} = y_{k-} = 0$ or $y_{k+}' = y_{k-}' = 0$, respectively. Finally, the conditions (\ref{3-fcoupl1})--(\ref{3-fcoupl4}) with $\alpha_{\mathrm{t}k} = 0$, $\beta_{\mathrm{t}k} \not = 0$, $\gamma_{\mathrm{t}k} = 2\frac{b_k^{1/2}+1}{b_k^{1/2}-1}$ change under the given transformation to
  $$
     y_{k+}' = -y_{k-}',\quad  y_{k+} + y_{k-} = \frac{\beta_{\mathrm{t}k}}{2} (b_k^{1/2}-1)^2 (-y_{k-}')\,,
  $$
while conditions $\alpha_{\mathrm{t}k} \not = 0$, $\beta_{\mathrm{t}k} = 0$, $\gamma_{\mathrm{t}k} = 2\frac{b_k^{1/2}+1}{b_k^{1/2}-1}$ change to
  $$
     y_{k+} =- y_{k-},\quad  y_{k+}' + y_{k-}' = -\frac{\alpha_{\mathrm{t}k}}{2} (b_k^{-1/2}-1)^2 y_{k-}\,.
  $$
 \end{lemma}

\section{Construction of the unitary equivalence}

With the above preliminaries, we are going to construct in this section the announced 
decomposition of $L_2 (\Gamma)$ into subspaces
of the radial functions and, subsequently, the equivalence of Hamiltonian
on a~tree graph to the orthogonal sum of halfline Hamiltonians. The construction extends the result of Appendix~A in Ref.~\onlinecite{HP} following the same line of reasoning.

By assumption $U_k$ is unitary, hence there are numbers $\theta_{k,j},\: j=1,\dots,b_k-1$, and a regular (in fact, unitary) matrix $W_k$ such that $U_k = W_k^{-1} D_k W_k$, where $D_k := \mathrm{diag}\,(\mathrm{e}^{i \theta_{k,1}},\dots,\mathrm{e}^{i \theta_{k,b_k-1}})$. For a given vertex $v$ of the $k$-th generation we can then define the operator $R_v$ on $H^{2}(\Gamma_{\succeq v})\ominus L^2_{0,\mathrm{\,rad}}(\Gamma_{\succeq v})$ which interchanges components on different subtrees emanating from this vertex,
 $$
   R_v :  \begin{pmatrix}f_1(x)\\ f_2(x)\\ \vdots\\ f_{b_k}(x)\end{pmatrix} \mapsto
   \begin{pmatrix}
   \sum_{j=1}^{b_k}(W_k \cdot V_k)_{1j}\,f_j (x)\\
   \sum_{j=1}^{b_k}(W_k \cdot V_k)_{2j}\,f_j (x)\\
   \vdots\\
   \sum_{j=1}^{b_k}(W_k \cdot V_k)_{(b_k-1)j}\,f_j (x)
   \end{pmatrix} \,;
 $$
here $f_j (x)$ is, of course, the wave function component on the $j$-th subtree.

To see how this transformation influences the coupling conditions, we start from the class of symmetric functions satisfying boundary conditions (\ref{3-fcoupl1})--(\ref{3-fcoupl4}),
 $$
   \mathrm{dom\,}\mathbf{H}_{o, \mathrm{rad}} =\mathrm{dom\,}\mathbf{H} \cap L^2_{0,\mathrm{\,rad}}(\Gamma_{\succeq o})\,.
 $$
Next we introduce for a~given vertex $v$ and $s = 1,\dots,b(v)-1$ the set
 \begin{multline*}
   \mathrm{dom\,}\mathbf{H}_{v s, \mathrm{rad}} =\{ f\in
    H^{2}(\Gamma_{\succeq v})\ominus L^2_{0,\mathrm{\,rad}}(\Gamma_{\succeq v})
  \,|\,\mathrm{supp\,}(R_v f)\subset \Gamma_{\succeq v, s},\\   (R_v f)_{vs+}' + (R_v f)_{vs+} \tan{\frac{\theta_{k s}}{2}}= 0,
    f\in L^2_{0,\mathrm{\,rad}}(\Gamma_{\succeq w}) \hbox{\,and satisfies\,(\ref{3-fcoupl1})--(\ref{3-fcoupl3}) for all }w\succeq v\}\,.
 \end{multline*}
where $\Gamma_{\succeq v, s}$ is the $s$-th subtree emanating from $v$.

 \begin{lemma}\label{r_v1}
 $f$ satisfies (\ref{3-fcoupl3}) iff $R_v f$ satisfies $(R_v f)_{vs+}' + (R_v f)_{vs+} \tan{\frac{\theta_{k s}}{2}}=0$ for all $s \in \{1,\dots , b(v)-1\}$.
 \end{lemma}
 \begin{proof}
 Substituting $U_k = W_k^{-1}D_k W_k$ into (\ref{3-fcoupl3}) and using the definition of the operator $R_v$ one obtains
 \begin{multline*}
   W_k^{-1} \left[\begin{pmatrix}
   \mathrm{e}^{i\theta_{k1}}&0&\dots&0\\
   0&\mathrm{e}^{i\theta_{k2}}&\dots&0\\
   \vdots&\vdots&\ddots&\vdots\\
   0&0&\dots&\mathrm{e}^{i\theta_{k(b_k-1)}}\end{pmatrix}
   -I\right]
   \begin{pmatrix}(R_v f)_{v1+}\\(R_v f)_{v2+}\\ \vdots\\(R_v f)_{v(b_k-1)+} \end{pmatrix}+
   \\ +
   i W_k^{-1} \left[\begin{pmatrix}
   \mathrm{e}^{i\theta_{k1}}&0&\dots&0\\
   0&\mathrm{e}^{i\theta_{k2}}&\dots&0\\
   \vdots&\vdots&\ddots&\vdots\\
   0&0&\dots&\mathrm{e}^{i\theta_{k(b_k-1)}}\end{pmatrix}
   +I\right]
   \begin{pmatrix}(R_v f)_{v1+}'\\(R_v f)_{v2+}'\\ \vdots\\(R_v f)_{v(b_k-1)+}' \end{pmatrix}=0
 \end{multline*}
which gives the desired formula.
\end{proof}

Now we can state the decomposition for the Hamiltonian domains.
 \begin{lemma} \label{3-lemdecompose}
One can decompose
 $$
   \mathrm{dom\,}\mathbf{H} = \mathrm{dom\,}\mathbf{H}_{o,\mathrm{rad}} \oplus\; \bigoplus_{\stackrel{v\in \Gamma}{v\not = o}}
    \bigoplus_{s=1}^{b(v)-1}\mathrm{dom\,}\mathbf{H}_{v s, \mathrm{rad}} \,.
 $$
 \end{lemma}
 \begin{proof}
By definition, functions from $\mathbf{H}_{o,\mathrm{rad}}$ and  $\mathbf{H}_{v s,\mathrm{rad}}$ satisfy conditions (\ref{3-fcoupl1})--(\ref{3-fcoupl2}) at every vertex $w \succeq v$. Since functions from $\mathbf{H}_{o,\mathrm{rad}}$ and  $\mathbf{H}_{v s,\mathrm{rad}}$ are radial, they do not influence condition (\ref{3-fcoupl3}) at any vertices $w \succ o$ and $w \succ v$, respectively. Finally, one infers from Lemma~\ref{r_v1} that condition (\ref{3-fcoupl3}) is preserved at $v$ in view of the relation $(R_v f)_{vs+}' + (R_v f)_{vs+} \tan{\frac{\theta_{k s}}{2}}=0$.
 \end{proof}

Let us introduce a family of simple quantum graphs which will be the building blocks of the decomposition. By $L_{ns}$ we denote a~halfline parametrized by $t\in [t_n, \infty)$ with coupling conditions of Lemma~\ref{3-lemma_th} at the points $t_k, \:k>n,$ and
the condition $y'+ \tan{\frac{\theta_{ns}}{2}}y=0$ at the endpoint $t_n$. Let further $L_0$ be a halfline $[0, \infty)$ with coupling condition (\ref{3-fcoupl4}) at $t = 0$. Now we define the operator $J_{vs}$ acting from $\mathrm{dom\,}\mathbf{H}_{v s, \mathrm{rad}}$ to $\mathrm{dom\,}H_{L_{ns}}$, i.e. to the set of halfline functions $f\in \bigoplus_{k=n}^{\infty}H^{2}(t_k,t_{k+1})$ satisfying the above described conditions, by
 $$
    J_{vs} f := \left. (R_v f)\right|_{e_n \subset \Gamma_{\succeq v,s}} \oplus\; \bigoplus_{k>n}\: (b_{n+1} \cdot\dots\cdot b_{k})^{1/2}  \left. (R_v f)\right|_{e_k \subset \Gamma_{\succeq v,s}}\,,
 $$
where $e_k \subset \Gamma_{\succeq v,s}$ is an edge emanating from a vertex of $k$-th generation.

 \begin{lemma}\label{3-lemma_j}
The operators $R_v$ and $J_{vs}$  are unitary.
 \end{lemma}
 \begin{proof}
Let $\tilde V_n$ be square $b_n \times b_n$ matrix which has the same entries in the first $b_n -1$ rows as $V_n$ and the $b_n$-tuple $(1/\sqrt{b_n}, \dots, 1/\sqrt{b_n})$ in the last row. Since $f \in \mathrm{dom\,}\mathbf{H}_{v s, \mathrm{rad}}$ does not contain a $L^2_{0,\mathrm{\,rad}}(\Gamma_{\succeq v})$ component, the relation $\|V_n (f_1, \dots, f_{b_n})^\mathrm{T}\|=\|\tilde V_n (f_1, \dots, f_{b_n})^\mathrm{T}\|$ obviously holds. Unitarity of the operator $R_v$ then follows from
  $$
   \| R_v f \|^2 = \| W_n V_n (f_1, \dots, f_{b_n})^\mathrm{T}\|^2 =
   \| \tilde V_n (f_1, \dots, f_{b_n})^\mathrm{T}\|^2 = \| f\|^2,
 $$
where we have employed unitarity of matrices $W_n$ and $\tilde V_n$.
Furthermore, for any $f \in \mathrm{dom\,}\mathbf{H}_{v s, \mathrm{rad}}$ we have the relation
 \begin{multline*}
   \|J_{vs} f \|^2_{L_{ns}} = \|R_v f\|^2_{e_n}+ \sum_{k>n}(b_{n+1} \cdot\dots\cdot b_{k})\,\|R_v f\|^2_{e_k}
   =\|R_v f\|^2_{\Gamma_{\succeq v,s}}=\|R_v f\|^2_{\Gamma_{\succeq v}}=\|f\|^2_{\Gamma_{\succeq v}}.
 \end{multline*}
Finally, the equality $\|R_v f\|^2_{\Gamma_{\succeq v,s}}=\|R_v f\|^2_{\Gamma_{\succeq v}}$ is due to $\mathrm{supp\,}(R_v f)\subset \Gamma_{\succeq v, s}$.
 \end{proof}
 \begin{lemma} \label{3-lemham}
Let $v$ be a~vertex belonging to the $n$-th generation.
The Hamiltonian $\mathbf{H}_{v s, \mathrm{rad}}$ is unitarily equivalent to $H_{L_{ns}}$, where $n=\mathrm{gen}\, v\:$ and
 $$
   H_{L_{ns}}:=-\frac{\mathrm{d}^2}{\mathrm{d}t^2} + V (t)
 $$
with the domain consisting of functions $f\in \bigoplus_{k=n}^{\infty}H^{2}(t_k,t_{k+1})$
satisfying the conditions of Lemma~\ref{3-lemma_th} at the points $t_k, k> n$ and $y'+ \tan{\frac{\theta_{ns}}{2}}y=0$ at $t_n$ with the potential $V(t) := V(|x|)$.
 \end{lemma}
 \begin{proof}
The claim follows easily from the construction described in Sec.~\ref{sec_mapping}, see Lemma~\ref{3-lemma_th}, in combination with Lemma~\ref{3-lemma_j}.
 \end{proof}

We can summarize the results of lemmata \ref{3-lemdecompose} and \ref{3-lemham} in following theorem.
 \begin{theorem} \label{3-thham}
The Hamiltonian $\mathbf{H}$ on a~radial tree graph $\Gamma$ is unitarily
equivalent to
 \begin{equation}
   \mathbf{H}\cong H_{L_0}\oplus\; \bigoplus_{n=1}^{\infty}\bigoplus_{s=1}^{b_n-1}(\oplus\, b_0\dots b_{n-1})H_{L_{ns}}\,.\label{decomp}
 \end{equation}
where $(\oplus\, m)H_{L_{ns}}$ is the $m$-tuple copy of the operator $H_{L_{ns}}$.
 \end{theorem}

In analogy with Ref.~\onlinecite{HP}, these results can be generalized also to so-called tree-like graphs (with the edges emanating from the vertices of the same generation replaced by the same compact graph).

Let us illustrate first a few steps of the construction in a simple situation.
\begin{example}
{\rm Consider a graph with $b_1 = 3$, which consist of the edge $(0,t_1)$ and three identical subgraphs $\Gamma_1$, $\Gamma_2$, $\Gamma_3$ (such as in Fig.~\ref{figtree}) connected to it by boundary conditions (\ref{3-fcoupl1})--(\ref{3-fcoupl3}). The $2 \times 2$ unitary matrix $U$ describing the coupling at the vertex of the first generation can be parametrized by four real numbers $\theta_1$, $\theta_2$, $\varphi$ and $r$,
 $$
    U = W^{-1} D W,\quad
    W = \begin{pmatrix}r \mathrm{e}^{i\varphi}& \sqrt{1-r^2}\,\mathrm{e}^{-i \varphi}\\ \sqrt{1-r^2}\,\mathrm{e}^{i \varphi}&-r \mathrm{e}^{-i\varphi} \end{pmatrix}\,,\quad
    D = \begin{pmatrix}\mathrm{e}^{i\theta_1} &0\\0&\mathrm{e}^{i\theta_2} \end{pmatrix}.
 $$
Let us choose
 $$
   V = \begin{pmatrix}\frac{1}{\sqrt{2}}&-\frac{1}{\sqrt{2}}&0\\\frac{1}{\sqrt{6}}&\frac{1}{\sqrt{6}}&-\frac{2}{\sqrt{6}}\end{pmatrix}\quad\Rightarrow\quad\tilde V = \begin{pmatrix}\frac{1}{\sqrt{2}}&-\frac{1}{\sqrt{2}}&0\\\frac{1}{\sqrt{6}}&\frac{1}{\sqrt{6}}&-\frac{2}{\sqrt{6}}\\ \frac{1}{\sqrt{3}}&\frac{1}{\sqrt{3}}&\frac{1}{\sqrt{3}}\end{pmatrix}\,.
 $$
to ensure that $\tilde V$ is unitary.
Then
 $$
   W V \!\!=\!\!\frac{1}{\sqrt{6}} \!
   \begin{pmatrix}
   \sqrt{3}r\mathrm{e}^{i\varphi}+\sqrt{1-r^2}\,\mathrm{e}^{-i\varphi}&-\sqrt{3}r\mathrm{e}^{i\varphi}+\sqrt{1-r^2}\,\mathrm{e}^{-i\varphi}&-2\sqrt{1-r^2}\,\mathrm{e}^{-i\varphi}\\
   \sqrt{3}\sqrt{1-r^2}\,\mathrm{e}^{i\varphi}-r\mathrm{e}^{-i\varphi}&-\sqrt{3}\sqrt{1-r^2}\,\mathrm{e}^{i\varphi}-r\mathrm{e}^{-i\varphi}&2r\mathrm{e}^{-i\varphi}
   \end{pmatrix}\,
 $$
and the operator which interchanges components $f_1(x)$, $f_2(x)$, $f_3(x)$ on $\Gamma_1$, $\Gamma_2$ and $\Gamma_3$ becomes
 $$
   R_1: \begin{pmatrix}f_1(x)\\ f_2(x)\\ f_3(x)\end{pmatrix} \to \begin{pmatrix}g_1(x)\\ g_2(x)\end{pmatrix}\,,
 $$
where
 \begin{multline*}
   g_1(x) = [\sqrt{3}r\mathrm{e}^{i\varphi}+\sqrt{1-r^2}\,\mathrm{e}^{-i\varphi}]f_1(x)+
\\
  +[-\sqrt{3}r\mathrm{e}^{i\varphi}+\sqrt{1-r^2}\,\mathrm{e}^{-i\varphi}] f_2(x) -2\sqrt{1-r^2}\,\mathrm{e}^{-i\varphi} f_3(x),
 \end{multline*}
 $$
   g_2(x) =[\sqrt{3}\sqrt{1-r^2}\,\mathrm{e}^{i\varphi}-r\mathrm{e}^{-i\varphi}] f_1(x)
  +[-\sqrt{3}\sqrt{1-r^2}\,\mathrm{e}^{i\varphi}-r\mathrm{e}^{-i\varphi}] f_2(x)+ 2r\mathrm{e}^{-i\varphi}f_3(x).
 $$
The boundary condition (\ref{3-fcoupl3}) for the vertex of the first generation then becomes
 $$
   (\mathrm{e}^{\theta_1}-1) g_1 (0) + i  (\mathrm{e}^{\theta_1}+1) g_1' (0) = 0\,,\quad  (\mathrm{e}^{\theta_2}-1) g_2 (0) + i  (\mathrm{e}^{\theta_2}+1) g_2' (0) = 0
 $$
which corresponds to the boundary conditions for the operators $H_{L_{11}}$ and $H_{L_{12}}$ at the halfline endpoint. The construction proceeds similarly for vertices of the next generations. }
\end{example}

\section{Absence of absolutely continuous spectra for halfline operators}

From now on we will suppose that the potential is absent, $V=0$. Our stated aim is to generalize the results of Breuer and Frank \cite{BF} to a much larger class of free Schr\"{o}dinger operators on trees. 
To be more specific, we are going to show that the result they proved for Laplacians  on trees with free (Kirchhoff) coupling remains valid for almost all coupling conditions which allow to perform the decomposition (\ref{decomp}). By ``almost all'' we mean here that possible exceptions correspond to a manifold of a lower dimension in the parameter space. 
We follow the same line of reasoning as in Ref.~\onlinecite{BF} showing that the absolutely continuous spectrum of halfline operators vanishes if the set of distances between the neighboring vertices contains a subsequence growing to infinity; the conclusion for tree graphs then follows from (\ref{decomp}).

We will consider one of the halfline operators $H_{L_n}= -\mathrm{d}^2/\mathrm{d}t^2$, for simplicity denoted by $H$, acting on functions which satisfy Dirichlet condition at $t = 0$ and conditions of Lemma~\ref{3-lemma_prevod} at the points $\{t_k\}_{k=1}^\infty$. For the sake of simplicity, we also drop the subscript h throughout this section, hence $\alpha$, $\beta$, $\gamma$, $a$, $d$, $c$ mean the corresponding halfline GPI coupling constants.

 \begin{lemma} \label{lemmaresol}
The resolvent of $H$ can be for $z \in \mathbb{C}\backslash [0,\infty)$ written as
 \begin{equation}
   (H-z)^{-1} =(H_0-z)^{-1} + (\mathrm{Tr\,} (H_0 - \bar z)^{-1})^* (T(z)+B)^{-1} \mathrm{Tr\,} (H_0 -z)^{-1}\,,\label{resolvent}
 \end{equation}
where $H_0$ acts as $-\mathrm{d}^2/\mathrm{d}t^2$ with the domain consisting of functions in $L^2(\mathbb{R}_+)$ which fulfil Dirichlet condition at $t_0$ and free conditions at the other vertices. The $2\times 2$ matrix operators $T(z)$ and $B$ are given by their entries
 $$
   T(z)_{nm} := \begin{pmatrix}
    \frac{1}{2ik}(\mathrm{e}^{ik|t_n - t_m|}-\mathrm{e}^{ik (t_n + t_m)})&
    \frac{1}{2}(\sigma_{mn}\mathrm{e}^{ik|t_n - t_m|}-\mathrm{e}^{ik (t_n + t_m)})\\[.3em]
    \frac{1}{2}(\sigma_{nm}\mathrm{e}^{ik|t_n - t_m|}-\mathrm{e}^{ik (t_n + t_m)})&
    -\frac{ik}{2}(\mathrm{e}^{ik|t_n - t_m|}+\mathrm{e}^{ik (t_n + t_m)})&
   \end{pmatrix}\,,
 $$
where $\sigma_{mn}:= \mathrm{sgn\,}(t_m - t_n)$, and
 $$
   B_{nm} := \delta_{nm} \frac{1}{\mathrm{det\,}\mathcal{A}_n}
   \begin{pmatrix} -\beta_n & -\gamma_n\\-\bar\gamma_n& \alpha_n \end{pmatrix}\,;
 $$
 the symbol $\mathrm{Tr}$ stands here for the trace operator from $L^2(\mathbb{R}_+)$ to $l(\mathbb{N},\mathbb{C}^2)$,
 $$
   (\mathrm{Tr\,} y)_n := \begin{pmatrix}y(t_n)\\y'(t_n)\end{pmatrix}\,.
 $$
 \end{lemma}
 \begin{proof}
The claim is a slight modification of Lemma~9 in Ref.~\onlinecite{BF} apart from the multiplication operator $B$. One can straghtforwardly check that 
 \begin{equation}
   \mathrm{Tr}_\pm (\mathrm{Tr}(H_0 -\bar z)^{-1})^* = -T(z)\pm \frac{1}{2}J \label{trpm}
 \end{equation}
with $J$ having the entries $J_{nm} = \delta_{nm}\begin{pmatrix}0&1\\-1&0\end{pmatrix}$ and $\mathrm{Tr}_\pm$ defined by $(\mathrm{Tr}_\pm y)_n = \begin{pmatrix}y(t_n\pm)\\y'(t_n\pm)\end{pmatrix}$ for all square integrable functions $y$ belonging to $W^{2,2}(t_n,t_{n+1})$ for each $n \geq 0$. Using the resolvent formula, the previous formula and the fact that functions in $\mathrm{Ran\,(H_0 -\zeta)^{-1}}$ and their first derivatives are continuous, i.e. $\mathrm{Tr}_\pm(H_0 -\zeta)^{-1} = \mathrm{Tr\,}(H_0 -\zeta)^{-1}$ one obtains
 $$
   T(z)-T(\zeta) = (\zeta - z) \mathrm{Tr\,} (H_0 -\zeta)^{-1} \left(\mathrm{Tr\,}(H_0-\bar z)^{-1}\right)^*\,.
 $$
As it follows from the result of Posilicano \cite{Po}, there is an operator $G$ with $(G-z)^{-1}$ equal to the \emph{rhs} of (\ref{resolvent}). Now we apply $\mathrm{Tr_\pm}(G-z)^{-1}$ to $(G-z)y$ for $y\in \mathrm{dom\,}G$. Denoting by $c = \mathrm{Tr\,}(H-z)^{-1}(G-z)y$ and $y_\pm = \mathrm{Tr}_\pm y$ and using $\mathrm{Tr}_\pm(H_0 -\zeta)^{-1} = \mathrm{Tr\,}(H_0 -\zeta)^{-1}$ and (\ref{trpm}) one obtains from (\ref{resolvent})
 $$
   y_\pm = c + \left(-T(z)\pm \frac{1}{2}J\right) \left(T(z)+B\right)^{-1} c = \left(B\pm \frac{1}{2}J\right)\left(T(z)+B\right)^{-1} c\,. 
 $$
The previous equation results to
 $$
   \begin{pmatrix}y(t_n+)\\y'(t_n+)\end{pmatrix} =
   \left(B_{nn} + \frac{1}{2}J_{nn}\right)\left(B_{nn} - \frac{1}{2}J_{nn}\right)^{-1}
   \begin{pmatrix}y(t_n-)\\y'(t_n-)\end{pmatrix}\,.
 $$
Since the coupling conditions (\ref{ycoupl1})--(\ref{ycoupl2}) can be rewritten in the form
 $$
   \begin{pmatrix}-\frac{\alpha_n}{2}&1-\frac{\gamma_n}{2}\\
    1+\frac{\bar\gamma_n}{2}& -\frac{\beta_n}{2}\end{pmatrix}
    \begin{pmatrix}y(t_n+)\\y'(t_n+)\end{pmatrix} =
    \begin{pmatrix}\frac{\alpha_n}{2}&1+\frac{\gamma_n}{2}\\
    1-\frac{\bar\gamma_n}{2}& \frac{\beta_n}{2}\end{pmatrix}
    \begin{pmatrix}y(t_n-)\\y'(t_n-)\end{pmatrix}\,,
 $$
the above expression of the operator $B$ can be easily verified.
 \end{proof}

We proceed by proving properties of the m-function defined as
 $$
   m_{\pm} (z,t) := \pm \frac{f_{\pm}'(z,t)}{f_{\pm}(z,t)},
 $$
where $f_{\pm}(z,t)$ are functions square integrable near $\pm \infty$, respectively, which solve the equation $-f''+zf = 0$ under the conditions (\ref{3-a1})--(\ref{3-a2}) at the point $t_n$.

 \begin{lemma} \label{lemmam}
  Let $T$ and $B$ be operators defined in Lemma~\ref{lemmaresol}. Then for the spectral parameter $z = k^2 \in \mathbb{C}\backslash [0,\infty)$, $\mathrm{Im\,}k>0$, the m-function at $t = 0$ is
 $$
   m_+ (k^2, 0) = ik + \sum_{n,m}\mathrm{e}^{ik(t_n +t_m)}\begin{pmatrix}1\\ik\end{pmatrix}^{\mathrm{T}} [(T(k^2)+B)^{-1})]_{n,m}\begin{pmatrix}1\\ik\end{pmatrix}\,.
 $$
 \end{lemma}
 \begin{proof}
 The argument is the same as in Corollary~11 in~Ref.~\onlinecite{BF}; Lemma~\ref{lemmaresol} in combination with the expression of the m-function from the Green's function
 $$
   m_+ (z, 0) = \left.\frac{\partial^2}{\partial t \partial u} (H -z)^{-1} (t,u)\right|_{{t,u} = (0,0)}\,.
 $$
 yields the result.
 \end{proof}

 \begin{lemma} \label{lemma_parametry_nenula}
Let the Hamiltonian $H$ satisfy coupling conditions (\ref{3-a1})--(\ref{3-a2}) with $\beta_n \ne 0$ for all $n\in\mathbb{N}$. Then its spectrum depends on the coupling parameters $a_n$, $b_n$, $|c_n|$ only, not on the phase of $c_n$.
 \end{lemma}
 \begin{proof}
Since  $\beta_n \ne 0$ it is more convenient to use parametrization (\ref{3-b}). Let $f = (f_1, f_2,f_3, \dots, f_n, \dots)$ be the solution of the problem with coupling conditions (\ref{3-b}) and parameters $a_n$, $b_n$, $|c_n|\,\mathrm{e}^{i\varphi_n}$ at the point $t_n$ and given a Robin condition at the root. The solution $f$ is unitarily equivalent to the solution
 $$
 \tilde f = (f_1, f_2\,\mathrm{e}^{-i\varphi_1},f_3\,\mathrm{e}^{-i(\varphi_1+\varphi_2)}, \dots, f_n\,\mathrm{e}^{-i\sum_{j=1}^{n-1}\varphi_j}, \dots)
 $$
 of the problem with coupling parameters $a_n$, $b_n$, $|c_n|$ at the point $t_n$ and the same Robin condition at the endpoint of the halfline.
 \end{proof}

Before proceeding with the proof that the m-function uniquely depends on parameters of the Hamiltonian let us formulate an analogue of a little bit technical Lemma~13 of Ref.~\onlinecite{BF}.
 \begin{lemma}\label{lemmatech}
Suppose that $t_1 > 0$, $\varepsilon = \mathrm{inf\,}_{n,m; n\not = m}|t_n -t_m|>0$ and $\mathrm{det\,}\mathcal{A}_1 \not = 0$.
If there exists $\delta > 0$ such that all $|\beta_n|>\delta$, then for large values of $\kappa$ we have
 \begin{multline}
   m_+ (-\kappa^2,0) + \kappa = -2\kappa\mathrm{e}^{-2\kappa t_1}\left[
   1 -2 d_1 \frac{1}{\kappa} + 2 (|c_1|^2 + d_1^2)\frac{1}{\kappa^2} \right.
\\
   \left. -2(a_1 |c_1|^2 + 2 |c_1|^2 d_1 + d_1^3)\frac{1}{\kappa^3} + \mathcal{O}\left(\frac{1}{\kappa^4}\right) \right]\,.
 \end{multline}
If all $\beta_n = 0$ then
 \begin{multline}
   m_+ (-\kappa^2,0) + \kappa = 2\kappa\mathrm{e}^{-2\kappa t_1}\left[
   \frac{4\mathrm{Re\,}\gamma_1}{4+|\gamma_1|^2}-\frac{2\alpha_1 (4 + |\gamma_1|^2+4\mathrm{Re\,}\gamma_1)}{(4+|\gamma_1|^2)^2}\frac{1}{\kappa}
\right. \\  \left.
   -\frac{4\alpha_1^2 (4 + |\gamma_1|^2+4\mathrm{Re\,}\gamma_1)}{(4+|\gamma_1|^2)^3}\frac{1}{\kappa^2}+ \mathcal{O}\left(\frac{1}{\kappa^3}\right)
   \right]\,.
 \end{multline}
  \end{lemma}

\noindent With the application to tree graphs in mind, we leave out the ``intermediate'' case when $\{\beta_n\}$ contains a subsequence which tends to zero.
 \begin{proof}
The first part is identical with the proof of Lemma~13 in Ref.~\onlinecite{BF}. Using the decomposition
 $$
   T(-\kappa^2) = T^0(-\kappa^2) + T^\mathrm{R}(-\kappa^2)\,,\quad \mathrm{where\ }T^0(-\kappa^2)_{nm} := \delta_{nm}\begin{pmatrix}-\frac{1}{2\kappa}&0\\0& \frac{\kappa}{2}\end{pmatrix}
 $$
and the bounds
 $$
   \|T^\mathrm{R}(-\kappa^2)_{nm}\|_{\hbox{\bbsmall C}\to \hbox{\bbsmall C}} \leq \left\{\begin{array}{ll}\mathrm{const\,}\kappa\,\mathrm{e}^{-2\kappa t_n}& \mathrm{for\;} n = m\\ \mathrm{const\,}\kappa\,\mathrm{e}^{-\kappa |t_n-t_m|}&\mathrm{for\;} n \not= m\end{array}\right.
 $$
one obtains the following estimate on the $l(\mathbb{N},\mathbb{C}^2)$ norm of $T^\mathrm{R}$ for large $\kappa$,
 $$
   \|T^\mathrm{R}(-\kappa^2)\| \leq \mathrm{const\,}\kappa (\mathrm{e}^{-2\kappa t_1}+\mathrm{e}^{-\kappa \varepsilon})\,.
 $$
The  operator $T^0(-\kappa^2)+B$ is under the given assumptions invertible. Let us check it first for $|\beta_n| > \delta$. The eigenvalues of $(T^{0}(-\kappa^2)+B)_{nn}$ are
 $$
   \lambda_1 = \frac{\kappa}{2} +\mathcal{O}\left(1\right), \quad  \lambda_2 = -\frac{\beta_n}{\mathrm{det\,}\mathcal{A}_n} + \mathcal{O}\left(\frac{1}{\kappa}\right)
 $$
being nonzero for large $\kappa$. On the other hand, in the case $\beta_n= 0$ we get
 $$
   \lambda_1 = \frac{\kappa}{2} +\mathcal{O}\left(1\right), \quad  \lambda_2 = -\frac{1}{2\kappa}\left(1+\frac{4}{|\gamma_n|^2}\right) + \mathcal{O}\left(\frac{1}{\kappa^2}\right)\,.
 $$
Hence the norm of the inverses of $T^0(-\kappa^2+B)$ and $T(-\kappa^2+B)$ is in both cases bounded above by a multiple of $\kappa$, which allows one to argue similarly as in the proof of Lemma~13 of Ref.~\onlinecite{BF},
 \begin{multline*}
   \|(T(-\kappa^2+B))^{-1} - (T^0(-\kappa^2+B))^{-1}\|
   \\[.3em]
    =\|(T(-\kappa^2+B))^{-1}T^\mathrm{R}(-\kappa^2) (T^0(-\kappa^2+B))^{-1}\| \leq \mathrm{const\,} \kappa^3 (\mathrm{e}^{-2\kappa t_1}+\mathrm{e}^{-\kappa \varepsilon})\,.
 \end{multline*}
Using Lemma~\ref{lemmam} and the fact that $[(T^{0}(-\kappa^2) +B)^{-1}]_{nn} = [(T^{0}(-\kappa^2)+B)_{nn}]^{-1}$ one can express $m_+ (\kappa,0) + \kappa$ as
 $$
   \sum_{n = 1}^{\infty} \mathrm{e}^{-2\kappa t_n} (1, \,- \kappa)\, (T^0(-\kappa^2)+B)_{nn}^{-1} \begin{pmatrix}1\\ - \kappa\end{pmatrix}
  +\mathcal{O}(\kappa^5 \mathrm{e}^{-2\kappa t_1}(\mathrm{e}^{-2\kappa t_1}+\mathrm{e}^{-\kappa \varepsilon}))\,.
 $$
Next we notice that the higher terms in the sum, $n\ge 2$, can be absorbed into the error term, and since
 $$
   (1, \,- \kappa)\, (T^0(-\kappa^2)+B)_{11}^{-1} \begin{pmatrix}1\\ - \kappa\end{pmatrix} = -\frac{4(\alpha_1-\beta_1\kappa^2-2\kappa \mathrm{Re\,}\gamma_1)}{{\mathrm{det\,}\mathcal{A}_1}
   +4+\frac{2}{\kappa}(\beta_1\kappa^2+\alpha_1)}
   =2\kappa\frac{\beta_1+\frac{2\mathrm{Re\,}\gamma_1}{\kappa}
   -\frac{\alpha_1}{\kappa^2}}{\beta_1+\frac{\mathrm{det\,}\mathcal{A}_1
   +4}{2\kappa}+\frac{\alpha_1}{\kappa^2}}
 $$
a straightforward computation yields the sought formul\ae.
 \end{proof}

With Lemma~\ref{lemma_parametry_nenula} in mind we define in the case that $\beta_n \not = 0$ for all $n\in\mathbb{N}$ the distance between a pair of full-line GPI Hamiltonians in analogy with Ref.~\onlinecite{R1},
 $$
   d(H^{(1)},H^{(2)}) := \sum_{m = 1}^\infty 2^{-m}\frac{\rho_m(H^{(1)},H^{(2)})}{1+\rho_m(H^{(1)},H^{(2)})}\,,
 $$
where
 $$
   \rho_m(H^{(1)},H^{(2)}):= \sum_{j=1}^{3} \left|\int_\mathbb{R} f_m (x) \,\mathrm{d}(\mu_j^{(1)}-\mu_j^{(2)})(x)\right|
 $$
with the measures $\mu_1^{(i)} := \sum_{n=1}^\infty a_n^{(i)}(t_n) \delta(t_n^{(i)})$, $\mu_2^{(i)} := \sum_{n = 1}^\infty d_n^{(i)}(t_n^{(i)}) \delta(t_n^{(i)})$, and $\mu_3^{(i)} := \sum_{n = 1}^\infty |c_n^{(i)}(t_n^{(i)})| \delta(t_n^{(i)})$; here $i \in \{1,2\}$ and $\{f_n: n\in\mathbb{N}\}$ is a compact subset of $C_\mathrm{c}(\mathbb{R})$ which is dense with respect to $\|.\|_\infty$. In contrast to Ref. \onlinecite{R1} we associate here three $\delta$ measures with each operator instead of one. In case when all the $\beta_n$'s vanish we define the distance similarly using two measures,
 $$
   \rho_m(H^{(1)},H^{(2)}):= \sum_{j=1}^{2} \left|\int_\mathbb{R} f_m (x) \,\mathrm{d}(\mu_j^{(1)}-\mu_j^{(2)})(x)\right|
 $$
with $\mu_1^{(i)} := \sum_{n=1}^\infty \frac{\mathrm{Re\,}\gamma_n^{(i)}}{|\gamma_n^{(i)}|^2+4}\, \delta(t_n^{(i)})$ and  $\mu_2^{(i)} := \sum_{n = 1}^\infty \frac{\alpha_n^{(i)}}{|\gamma_n^{(i)}|^2+4}\, \delta(t_n^{(i)})$.

 \begin{theorem}\label{thmuniq}
Suppose that the m-functions of two GPI Hamiltonians $H^{(1)}$ and $H^{(2)}$ satisfy $m_+^{(1)}(z,t) = m_+^{(2)}(z,t)$ for some $t< \mathrm{min\,}(t_1^{(1)},t_1^{(2)})$ and for all $z\in\mathbb{C}$. Furthermore, assume that neither $H^{(1)}$ nor $H^{(2)}$ contains a GPI with separating coupling conditions (corresponding to $\mathrm{det\,}\mathcal{A} = 4$ and $\mathrm{Im\,}\gamma = 0$) and that all the coupling conditions fulfil the assumptions of Lemma~\ref{lemmatech}. Then $d(H^{(1)},H^{(2)}) = 0$ which specifically means
\begin{itemize}
\item[(a)]
for $|\beta_n|>\delta,\: \forall n\in\mathbb{N}$: $H^{(1)}$ equals $H^{(2)}$ up to the equivalence relation given by a phase change of the coefficients $c_n$.
\item[(b)]
for $\beta_n = 0,\: \forall n\in\mathbb{N}$: $H^{(1)}$ equals $H^{(2)}$ up to possible coefficient transformations which satisfy $\frac{\mathrm{Re\,}\gamma_n^{(1)}}{|\gamma_n^{(1)}|^2+4} =\frac{\mathrm{Re\,}\gamma_n^{(2)}}{|\gamma_n^{(2)}|^2+4}$ and $\frac{\alpha_n^{(1)}}{|\gamma_n^{(1)}|^2+4} = \frac{\alpha_n^{(2)}}{|\gamma_n^{(2)}|^2+4}$.
\end{itemize}
 \end{theorem}
 \begin{proof}
The argument is similar to that in the proof of Proposition~12 in Ref.~\onlinecite{BF}. The expressions for large $\kappa$ limit in both cases considered in Lemma~\ref{lemmatech} determine $t_1$ and all the coupling parameters at $t_1$. With the exception of the separating conditions case one can uniquely solve the equation $-y'' = z y$ on $(0,s)$, $s>t_1$ and hence to obtain $m_+^{(1)}(s,z) = m_+^{(2)}(s,z)$.
 \end{proof}

In order to formulate an analogue of Remling theorem suitable for our purpose we introduce --- using a self-explanatory notion --- the set of right-limits of a halfline operator $H(\{t_n\}_{n = 1}^{\infty},\{\mathcal{A}_n\}_{n=1}^\infty)$ as the set $\omega(H(\{t_n\}_{n = 1}^{\infty},\{\mathcal{A}_n\}_{n=1}^\infty))$ consisting of those full-line GPI Hamiltonians $\hat H$ for which there is  a strictly increasing sequence $\{s_m\},\: s_m\to\infty\,$, such that
 $$
   d(H'(\{t_n+s_m\}_{n = 1}^{\infty},\{\mathcal{A}_n\}_{n=1}^\infty),\hat H) \to 0
 $$
holds as $m\to\infty$. $H'(\{t_n+s_m\}_{n = 1}^{\infty},\{\mathcal{A}_n\}_{n=1}^\infty)$ stands for a full-line operator which acts freely on $(-\infty,t_1)$ and satisfies the same coupling conditions at $t_n$, $n\in\mathbb{N}$ as $H(\{t_n+s_m\}_{n = 1}^{\infty},\{\mathcal{A}_n\}_{n=1}^\infty)$.

 \begin{theorem}\label{remling}
Let $H(\{t_n\}_{n = 1}^{\infty},\{\mathcal{A}_n\}_{n=1}^\infty)$ be a GPI Hamiltonian without separating coupling conditions. Then any right limit $\hat H \in \omega(H(\{t_n\}_{n = 1}^{\infty},\{\mathcal{A}_n\}_{n=1}^\infty))$ is reflectionless on $\Sigma_{\mathrm{ac}}(H(\{t_n\}_{n = 1}^{\infty},\{\mathcal{A}_n\}_{n=1}^\infty))$, in other words, the relation $\hat m_+(E +i 0,t) = -\bar{\hat m}_-(E +i 0,t)$ holds for all $t\in \mathbb{R}\backslash\{t_n\}$ and almost every energy value $E\in\Sigma_{\mathrm{ac}}(H(\{t_n\}_{n = 1}^{\infty},\{\mathcal{A}_n\}_{n=1}^\infty))$.
 \end{theorem}
 \begin{proof}
The proof works in the same way as in Theorem~16 of Ref.~\onlinecite{BF}.
Omitting for simplicity the subscript $n$, we can rewrite the coupling conditions (\ref{3-a1})--(\ref{3-a2}) in the form
 $$
   \begin{pmatrix}f_+ \\ f_+'\end{pmatrix} =
   \frac{1}{4-\mathrm{det\,}\mathcal{A}-4i \mathrm{Im\,}\gamma}\,\begin{pmatrix}4+\mathrm{det\,}\mathcal{A}-4 \mathrm{Re\,}\gamma& 4\beta \\4\alpha & 4+\mathrm{det\,}\mathcal{A}+4 \mathrm{Re\,}\gamma \end{pmatrix}\begin{pmatrix}f_- \\ f_-'\end{pmatrix}\,.
 $$
It is straightforward to check that in the non-separating case we have
 $$
   f_- \bar g_-' - f_-' \bar g_- =f_+ \bar g_+' - f_+' \bar g_+ \,,\quad |f_- 
   g_-' - f_-' g_- | = |f_+  g_+' - f_+' g_+ |\,,
 $$
and since Green's formula
 $$
   \int_a^b (-f''(t))\bar g (t)\,\mathrm{d}t-\int_a^b f(t)(-\bar g''(t))\,\mathrm{d}t  = W (f,\bar g) (b) - W (f,\bar g) (a)
 $$
holds in our case, one can employ Weyl nested disc construction (see, e.g., Ref.~\onlinecite{CL}) to prove that $\lim_{j\to \infty}m_+(z,s_j)= \hat m_+(z,0)$. To be more specific, if one defines solutions $u$, $v$ satisfying the initial coupling conditions 
 $$
   u(0) = 1, \quad u'(0) = 0,\quad v(0) = 0, \quad v'(0) = 1
 $$
then from the definition of the m-function follows $f(x) = u(x)+m(z,0) v(x) \in L^2(0,\infty)$. Any Robin coupling condition at $x = b$
 $$
   \cos{\omega} f(b) + \sin{\omega} f'(b) = 0\,,\quad \omega\in[0,\pi)
 $$
leads to a M\"{o}bius transformation
 $$
   m(z,0) = -\frac{\cot{\omega} \,u(b)+ u'(b)}{\cot{\omega}\, v(b)+ v'(b)}.
 $$
One can straightforwardly show that the image of the real axis under this transformation is the circle with the center $W(u,\bar v)(b)/W(v,\bar v)(b)$ and the radius $|W(u, v)(b)|/|W(v,\bar v)(b)|$. Since in the limit circle case there is no absolutely continuous spectrum, one can assume the limit point case and establish the convergence $\lim_{j\to \infty}m_+(z,s_j)=\hat m_+(z,0)$. In a similar way, one can prove $\lim_{j\to \infty}-v'(z,s_j)/v(z,s_j) = \hat m_-(z,0)$. The claim now follows from Theorem~1 in Ref.~\onlinecite{BP}.

 \end{proof}

 \begin{theorem}\label{th-halfline}
Let $H$ be the halfline GPI Hamiltonian with Dirichlet condition at $t = 0$ and coupling conditions (\ref{3-a1}) and (\ref{3-a2}) at the points $t = t_n$. Let the coupling constants at each vertex $t_n$ satisfy the assumptions of Lemma~\ref{lemmatech} and let there exist $N\in \mathbb{N}$, $K\in(0,\infty)$ and $\delta>0$ such that for all $n>N$  one of the following conditions holds: either
\begin{enumerate}
\item[(a)] $|\beta_n|>\delta>0$ and $|c_n|>\delta>0$, or
\item[(b)] $\beta_n = 0$, $|\gamma_n|<K$, and at least one of the following conditions is valid for all $n>N$: $\mathrm{Re\,}\gamma_n >\delta$ or $\mathrm{Re\,}\gamma_n <-\delta$ or $\alpha_n >\delta$ or $\alpha_n <-\delta$.
\end{enumerate}
Suppose that the number of GPI's described by separating conditions is at most finite. Let $\varepsilon = \mathrm{inf\,}_{n,m; n\not = m}|t_n -t_m|>0$. If $\mathrm{lim\,sup}_{n\to \infty}(t_{n+1}-t_n) = \infty$, the absolutely continuous spectrum of H is empty.
 \end{theorem}
 \begin{proof}
First, notice that the result is insensitive to the presence of a finite number of separating conditions (i.e., such that $\mathrm{det\,} \mathcal{A} = 4$ and $\mathrm{Im\,}\gamma = 0$). Since a change of boundary conditions is a rank-one perturbation of the resolvent which does affect the \emph{ac} spectrum, we may replace the rightmost among such conditions by Dirichlet and consider the halfline to the right of this point. The left out part corresponds to a finite interval, and therefore it does not contribute to the \emph{ac} spectrum.

The rest of the argument proceeds in analogy with the proof of Theorem~6 in Ref.~\onlinecite{BF}. Choosing a subsequence $\{s_j\}$ of the sequence $\{t_j -\varepsilon/2\}$ and mimicking the reasoning from Ref.~\onlinecite{BF} one can conclude that there are measures $\mu_{i}(t+s_j)$ which converge $\ast$-weakly to some $\hat\mu_{i}(t)$ as $j\to \infty$. Moreover, since $\mu_3(t_n)$ in the case (a) and at least one of the sequences $\pm \mu_1(t_n)$, $\pm \mu_2(t_n)$ is bounded from below by $\delta$, at least one of the measures $\hat\mu_i$ satisfies $\hat\mu_i(0,\infty)\not=0$. On the other hand, since $\mathrm{lim\,sup}_{n\to \infty}(t_{n+1}-t_n) = \infty$ we have $\hat\mu_i(-\infty,0) = 0$. Thus the full-line operator corresponding to $H$ has a right limit $\hat H$ which acts as the free operator on $(-\infty,0]$ (this implies, in particular, $\hat m_- (k^2+i0)= i k$) and it is nontrivial on $(0,\infty)$.

Suppose that $\Sigma_{\mathrm{ac}}(H)$ has a positive Lebesgue measure, then from Theorem~\ref{remling} we get $\hat m_+(k^2+i0)=-\bar{\hat m}_-(k^2+i0)=i k$ for all $k^2\in \Sigma_{\mathrm{ac}}(H)$ and $t\not = t_n$. Since the m-function is a Herglotz function, it is uniquely determined by its values on a set of positive Lebesgue measure. From Theorem~\ref{thmuniq} we conclude that the m-function of $\hat H$ corresponds to the free Hamiltonian. Noting that under the assumptions given above no coefficient transformation indicated in Theorem~\ref{thmuniq} can relate the free Hamiltonian and $\hat H$, we arrive thus at a contradiction.
 \end{proof}

\section{Absence of absolutely continuous spectra for trees}
With the unitary equivalence (\ref{decomp}) in mind, the application of the previous section results to radial tree graphs is simple. For notational convenience we will first write down several conditions needed in the following:
\begin{eqnarray}
&&\mathrm{det}\mathcal{A}_{\mathrm{t}n} (\sqrt{b_k}-1)+4 (1-b_n)\mathrm{\,Re\,}\gamma_{\mathrm{t}n}+ 4 (1+\sqrt{b_n})\not = 0 \phantom{AAAAAAAA}  \label{tri} \\
&&\frac{1}{K}<\big|4-2\sqrt{b_n}(\mathrm{det}\mathcal{A}_{\mathrm{t}n}-4)+ \mathrm{det}\mathcal{A}_{\mathrm{t}n}\nonumber \\ &&\qquad + b_n(4+\mathrm{det}\mathcal{A}_{\mathrm{t}n}-4\mathrm{Re\,}\gamma_{\mathrm{t}n})
+4\mathrm{Re\,}\gamma_{\mathrm{t}n}\big|<K \label{ctyri} \\
&& \frac{1}{K}<4 b_n \mathrm{det}\mathcal{A}_{\mathrm{t}n} + (1-b_n)[(4+\mathrm{det}\mathcal{A}_{\mathrm{t}n}+ 4\mathrm{Re\,}\gamma_{\mathrm{t}n})^2 \nonumber \\ &&\qquad -b_n(4+\mathrm{det}\mathcal{A}_{\mathrm{t}n}-4 \mathrm{Re\,}\gamma_{\mathrm{t}n})^2]<K \label{pet} \\
&& \frac{b_n^{1/2}}{|\beta_{\mathrm{t}n}|}\sqrt{(-4+\mathrm{det}\mathcal{A}_{\mathrm{t}n})^2
+(4\,\mathrm{Im\,}\gamma_{\mathrm{t}n})^2}>1/K \label{sest}
\end{eqnarray}
Using them we are able to state our main result.
 \begin{theorem} Let $\mathbf{H}$ be the Hamiltonian acting as $-\mathrm{d}^2/\mathrm{d} x^2$ on a radial tree graph with branching numbers $b_n$ and the domain consisting of all functions $f\in\bigoplus_{e\in \Gamma} H^{2}(e)$ satisfying the coupling conditions (\ref{3-fcoupl1})--(\ref{3-fcoupl4}) at $t_n$, $n\in\mathbb{N}$, among which the number of separating ones is at most finite. Suppose that there are $K \in (0,\infty)$ and $N \in\mathbb{N}$ such that for all $n>N$ the following conditions hold:
\begin{enumerate}
\item[(i)] $\mathrm{lim\,sup}_{n\to \infty}(t_{n+1}-t_n) = \infty$,
\item[(ii)] $\mathrm{inf}_{m,n}(t_{m}-t_n) > 0$,
\item[(iii)] either $\mathrm{\,Im\,}\gamma_{\mathrm{t}n} \not = 0$, or  both $\mathrm{det}\mathcal{A}_{\mathrm{t}n} \not =4$ and condition (\ref{tri}) are valid,
\item[(iv)] conditions (\ref{ctyri}) and (\ref{pet}) hold,
\item[(v)] finally, one of the following conditions holds:
  \begin{enumerate}
  \item[(a)] $b_n |\beta_{\mathrm{t}n}|>\frac{1}{K}$ and (\ref{sest}) is valid for all $n>N$,
  \item[(b)] $\beta_{\mathrm{t}n} = 0$, and either the right-hand side of (\ref{alphah}) is larger than $1/K$ for all $n>N$ or smaller than $-1/K$ for all $n>N$, or the \emph{rhs} of (\ref{betah}) is larger than $1/K$ for all $n>N$ or smaller than $-1/K$ for all $n>N$.
  \end{enumerate}
\end{enumerate}
Then the absolutely continuous spectrum of $\mathbf{H}$ is empty.
 \end{theorem}
 \begin{proof}
The claim follows from Theorems~\ref{3-thham} and \ref{th-halfline} in combination with the fact that absolutely continuous spectrum is not affected by a change of coupling conditions at a finite number of vertices. The assumptions can be obtained by a direct rephrasing of Lemmata~\ref{3-lemma_prevod} and \ref{3-lemma_th}. The assumptions (i) and (ii) constrain the variation edge lengths, (iii) excludes (an infinite number of) separating conditions, (iv) restricts denominators in Lemmata~\ref{3-lemma_th} and $\mathrm{det\,}\mathcal{A}_{\mathrm{h}n}$, respectively. Finally, (v) ensures that the assumptions of the previous theorem are satisfied.
 \end{proof}

One should keep in mind, however, that although the above result holds for quite a large family of coupling conditions, there are cases of trees which are sparse, $\mathrm{lim\,sup}_{n\to \infty}(t_{n+1}-t_n) = \infty$, but all the same their spectrum contains an absolutely continuous part or even is purely absolutely continuous. The most obvious one looks as follows.
 \begin{example}
{\rm Consider trees for which there is an $N$ that for all $n\in \mathbb{N},\, n \geq N$ one has $\alpha_{\mathrm{t}n} =\beta_{\mathrm{t}n}=0$, while $\gamma_{\mathrm{t}n} = 2\frac{b_n^{1/2}-1}{b_n^{1/2}+1}$. Then the spectrum of corresponding Hamiltonian contains an absolutely continuous part. In particular, if $N = 1$, then the spectrum is purely absolutely continuous. These claims are easy to check. As one can see from Lemma~\ref{3-lemma_th}, all halfline components in the decomposition (\ref{decomp}) act the right of the point $t_n$ as the free Hamiltonian, $\alpha_{\mathrm{h}n} =\beta_{\mathrm{h}n}=\gamma_{\mathrm{h}n} =0$. Consequently, the absolutely continuous spectrum of each component contains the interval $[0,\infty)$. If $N =1$, the tree Hamiltonian decomposes by (\ref{decomp}) to an infinite family of free halfline Hamiltonian copies with Dirichlet condition at the root and one with Robin condition (\ref{3-fcoupl4}). Note that these conclusions are not sensitive to the distribution of the points $\{t_n\}$, in particular, they hold for sparse trees considered here. }
 \end{example}

The last result allows for various modifications. For instance, one can keep $\alpha_{\mathrm{t}n}=\beta_{\mathrm{t}n}=0$ and change the above used parameter to $\gamma_{\mathrm{t}n} = 2\frac{b_n^{1/2}+1} {b_n^{1/2}-1}$ at some or all vertices. The claims are preserved, since such a coupling corresponds in view of Lemma~\ref{3-lemma_th} to the conditions $y_+ =- y_-,\,y_+' = -y_-'$ on the halfline, which are unitarily equivalent to the free coupling.

What is more important the decomposition (\ref{3-fcoupl4}) was derived not only for free operators. If we thus take a Hamiltonian with the coupling conditions of the above example acting as $-\mathrm{d}^2/\mathrm{d} x^2 + V(|x|)$ with a potential $V\in L^2(\mathbb{R}_+)$ then by the known result of Ref.~\onlinecite{DK} the claims we made remain valid.

\section*{Acknowledgments}
The research was supported by the Czech Ministry of Education,
Youth and Sports within the project LC06002. We are grateful to Rupert Frank
for useful comments.

\end{document}